\documentclass[11pt]{article}
\usepackage[a4paper, margin=1in]{geometry}
\usepackage{amsmath, amssymb}
\usepackage{graphicx}
\usepackage{titlesec}
\usepackage{enumitem}
\usepackage{booktabs}
\usepackage{float}
\usepackage{tikz}
\usetikzlibrary{arrows.meta,positioning}
\usepackage{url}  
\usepackage{hyperref}  
\usepackage[numbers,square]{natbib}  

\titleformat{\section}
{\normalfont\Large\bfseries}{\thesection.}{1em}{}

\begin{document}

\begin{center}
	\Large \textbf{Measuring Ransomware Lateral Movement Susceptibility via Privilege-Weighted Adjacency Matrix Exponentiation} \\
	\vspace{0.3cm}
	\normalsize
	\textbf{Satyam Tyagi}$^{1}$ \quad \textbf{Ganesh Murugesan}$^{1}$ \\
	\vspace{0.2cm}
	$^{1}$\textit{ColorTokens Inc., San Jose, CA, USA} \\
	\vspace{0.2cm}
	$^{1}$\texttt{satyam.tyagi@colortokens.com} \,\textbullet\, ORCID: \texttt{0009-0001-0137-2338} \\
	$^{1}$\texttt{ganesh.murugesan@colortokens.com} \,\textbullet\, ORCID: \texttt{0009-0001-4108-4752}
\end{center}

\vspace{0.4cm}
\textbf{Abstract:}
Ransomware impact hinges on how easily an intruder can move laterally and spread to the maximum number of assets.
We present a graph-theoretic formulation that casts lateral movement as a path-closure problem over a probability semiring to measure lateral-movement susceptibility and estimate blast radius.
We build a directed multigraph where vertices represent assets and edges represent reachable services (e.g., RDP/SSH)
between them. We model lateral movement as a probabilistic process using a pivot potential factor $\pi(s)$
for each service, with step successes composed via a probabilistic path operator \( \otimes \) and alternative paths aggregated via a probabilistic union \( \oplus \) (noisy-OR).
This yields a monotone fixed-point (iterative) computation of a $K$-hop compromise probability matrix that captures
how compromise propagates through the network. Metrics derived from this model include: (1) Lateral-Movement
Susceptibility (LMS$_K$): the average probability of a successful lateral movement between any two assets (0-1 scale); and (2)
Blast-Radius Estimate (BRE$_K$): the expected percentage of assets compromised in an average attack scenario.
Interactive services (SSH 22, RDP 3389) receive higher $\pi(s)$ than app-only ports (MySQL 3306, MSSQL 1433),
which seldom enable pivoting without an RCE. Across anonymized enterprise snapshots, pruning high-$\pi(s)$
edges yields the largest LMS$_K$/BRE$_K$ drop, aligning with CISA guidance, MITRE ATT\&CK (TA0008: Lateral Movement),
and NIST SP~800-207. The framework evaluates (micro)segmentation and helps prioritize controls that reduce
lateral-movement susceptibility and shrink blast radius.

\vspace{0.3cm}

\vspace{0.3cm}

\section{Introduction}\label{sec:intro}

Ransomware impact is determined less by where an adversary first lands and more by how easily they can \emph{move laterally} to reach high-value assets and propagate encryption. Classic reachability (Boolean adjacency, $k$-hop closure) reveals where paths exist, but it can \emph{overstate} risk by treating an SSH or RDP foothold as equivalent to a low-privilege application socket. In practice, interactive remote services (e.g., RDP/SSH) enable immediate pivoting; app-only ports typically demand additional exploits. Our goal is a probability-weighted alternative that measures the \emph{ease of movement}—and the expected blast radius—using only topology plus coarse service-class information.

We introduce a privilege- and account-aware adjacency model, casting lateral movement as a \emph{path-closure problem over a probability semiring}.
The model composes per-service ``pivot potentials'' $\pi(s)\!\in\![0,1]$ with a \emph{probabilistic path operator} $\otimes$ for sequential step composition and a \emph{probabilistic union} $\oplus$ across alternative routes (noisy-OR).
Iterating this algebraic path-closure yields (i) a $K$-hop (or $\varepsilon$-fixed-point) \emph{compromise probability matrix} and (ii) two deployment metrics: \textbf{LMS$_K$} (lateral-movement susceptibility; mean off-diagonal) and \textbf{BRE$_K$} (blast-radius estimate; mean including the diagonal).

\paragraph{CVE-free lateral movement via valid accounts.}
Once an adversary holds \emph{valid credentials} and an \emph{interactive remote service} is exposed (RDP/SSH), lateral movement does \emph{not} require unpatched vulnerabilities: adversaries log into remote services and act as the authenticated user \citep{mitre-ta0008,mitre-t1021,mitre-t1078}. Patch status is orthogonal to pivotability when privileged landing is possible; hence our operators weight \emph{service exposure and privilege class}, not CVE likelihoods.

\paragraph{Contributions.}
\begin{itemize}
	\item \textbf{Privilege- and account-aware path closure.} We formalize lateral movement over $[0,1]$ with \emph{probabilistic path composition} ($\otimes$) and \emph{probabilistic union} ($\oplus$), explicitly capturing valid-account pivots on interactive services \citep{mitre-t1021}.
	\item \textbf{Service-class lifting.} We encode exact ``land-then-pivot'' semantics by lifting states to $(\text{node},\text{class})$, preserving first-hop reach while applying class-conditioned pivot probabilities thereafter.
	\item \textbf{Actionable metrics.} We define \textbf{LMS$_K$} and \textbf{BRE$_K$} for policy comparison and segmentation planning.
	\item \textbf{Theoretical bounds.} We show that our probabilistic model $P_K$ is upper-bounded (entrywise) by the $K$-hop Boolean reachability $R_K$ and establish a single-path product lower bound (cf. Section~\ref{sec:formal-properties}).
	\item \textbf{Alignment with guidance.} Our empirical reductions in LMS$_K$/BRE$_K$ under RDP/SSH constraints match Zero Trust and ransomware guidance to limit remote desktop services and constrain lateral movement \citep{nist-800-207,cisa-stopransomware-2025,cisa-rdp-disable,cisa-rdp-restrict}.
\end{itemize}

\paragraph{Relation to Zero Trust and segmentation.}
Zero Trust (NIST SP~800-207) calls for minimizing implicit trust and continuously evaluating access in context \citep{nist-800-207}. Our metrics quantify the \emph{residual ease of movement} under candidate policies, complementing microsegmentation by highlighting edges whose removal yields the steepest probability-mass drop.

\paragraph{Organization.}
Section~\ref{sec:related} reviews the literature; Section~\ref{sec:graph-foundations} develops the graph-theoretic underpinnings; Section~\ref{sec:proposed-model} formalizes the matrices, operators ($\oplus$ and $\otimes$), and computation for the proposed model; Section~\ref{sec:quantifying-metrics} defines the metrics LMS$_K$ and BRE$_K$; Section~\ref{sec:formal-properties} presents formal properties and bounds; Section~\ref{sec:examples} provides examples and evidence; and Section~\ref{sec:conclusion} concludes with discussion and future work.

\section{Background and Related Work}\label{sec:related}

\textbf{Account-based pivots vs.\ vulnerability-driven traversal.}
Beyond exploit-based movement (e.g., T1210) \citep{mitre-t1210}, ATT\&CK catalogs techniques enabling lateral movement without software flaws: \emph{Valid Accounts} (T1078) and \emph{Remote Services} (T1021, including RDP/SSH) describe abusing legitimate credentials to authenticate and operate on remote hosts \citep{mitre-ta0008,mitre-t1021,mitre-t1078,mitre-t1550}. Our formulation treats interactive services as \emph{high-pivot} edges when paired with account access, independent of CVE posture.

\textbf{Attack graphs and probabilistic security metrics.}
Classical attack graphs model preconditions/exploits and enumerate paths to goals \citep{phillips-swiler-1998,sheyner-2002}. Later work augments graphs with probabilities and costs. We deliberately \emph{do not} encode per-CVE logic; instead we treat reachable services as coarse pivot opportunities whose probabilities compose via a \emph{probabilistic path operator} ($\otimes$) with \emph{probabilistic union} ($\oplus$) across alternatives, trading granularity for deployability on routine network snapshots.

\textbf{Algebraic path problems and operator view.}
Many path computations admit uniform treatment via algebraic path frameworks replacing $(+,\times)$ by problem-specific $(\oplus,\otimes)$. We adopt a \emph{probabilistic path operator} $\otimes$ for per-step composition and a \emph{probabilistic union} $\oplus$ for alternatives (noisy-OR), yielding a monotone iteration toward a least fixed point; this is analogous to algebraic path problems—we do not assume a semiring ($\oplus$ and $\otimes$ are non-idempotent, and distributivity holds only as an inequality) \citep{mohri-2002,sanmartin-2022}.

\textbf{Probabilistic union (noisy-OR) and causal independence.}
Our probabilistic union operator corresponds to the standard noisy-OR from probabilistic graphical models, aggregating independent causal mechanisms with low elicitation burden \citep{heckerman-1994}. Generalizations (e.g., Noisy-MAX) clarify when such factorizations are appropriate.

\textbf{Epidemic spreading analogies.}
Probabilistic composition along edges with probabilistic union across parallel paths is mathematically analogous to network contagion; classic results connect topology to spread \citep{newman-2002}. Our aim is \emph{risk measurement} (expected reach/blast), not epidemic prediction, but the analogy explains the outsized effect of a small set of high-pivot edges.

\textbf{AD attack-path tooling.}
In enterprise Windows domains, BloodHound operationalizes graph-based lateral-movement analysis at scale \citep{bloodhound-docs}. We follow similar instincts (directed, typed edges) but compute a \emph{probability-weighted} closure and provide system-level metrics for control evaluation.

\textbf{Zero Trust and ransomware guidance.}
NIST SP~800-207 sets the architectural principle; ransomware advisories emphasize limiting RDP/remote services to reduce both initial access and lateral movement \citep{nist-800-207,cisa-stopransomware-2025,cisa-rdp-disable,cisa-rdp-restrict}.

\section{Graph-theoretic underpinnings}\label{sec:graph-foundations}

\subsection{Adjacency matrices and notation}
Let $A$ be the (directed) adjacency matrix of the network; $A_{ij}{=}1$ iff a policy-permitted, actively listening service on $j$ is reachable from $i$, and $A_{ii}{=}0$; later we use a pivot-potential matrix $W\!\in[0,1]^{n\times n}$ with $W\le A$ (elementwise).

\begin{figure}[H]
	\centering
	\begin{minipage}{0.35\linewidth}
		\begin{tikzpicture}[>=Stealth, scale=0.65]
			\coordinate (i) at (0,0);
			\coordinate (j) at (2,1);
			\coordinate (k) at (2,-1);
			\coordinate (l) at (4,0);

			\node (ni) [draw, circle, inner sep=1.6pt] at (i) {$i$};
			\node (nj) [draw, circle, inner sep=1.6pt] at (j) {$j$};
			\node (nk) [draw, circle, inner sep=1.6pt] at (k) {$k$};
			\node (nl) [draw, circle, inner sep=1.6pt] at (l) {$\ell$};

			\draw[->,bend left=12]  (ni) to (nj);
			\draw[->,bend right=12] (ni) to (nk);
			\draw[->,bend left=12]  (nj) to (nk);
			\draw[->,bend left=14]  (nj) to (nl);
			\draw[->,bend right=14] (nk) to (nl);
		\end{tikzpicture}
	\end{minipage}\hspace{1mm}%
	\begin{minipage}{0.22\linewidth}
		{\setlength{\abovedisplayskip}{6pt}%
			\setlength{\belowdisplayskip}{6pt}%
			\setlength{\arraycolsep}{3pt}%
			\[
				A=\begin{bmatrix}
					0 & 1 & 1 & 0 \\
					0 & 0 & 1 & 1 \\
					0 & 0 & 0 & 1 \\
					0 & 0 & 0 & 0
				\end{bmatrix}
			\]
		}
	\end{minipage}
	\caption{Directed graph and its adjacency matrix $A$.}
	\label{fig:step1-A}
\end{figure}

\subsection{$K$-hop \emph{path} reachability (Boolean closure)}
Let $A\in\{0,1\}^{n\times n}$ be the adjacency matrix ($A_{ij}{=}1$ iff there is an edge $i{\to}j$).
Over the Boolean semiring $(\lor,\land)$, define the Boolean matrix product
$(X\diamond Y)_{ij}:=\bigvee_{\ell=1}^n X_{i\ell}\land Y_{\ell j}$,
Boolean powers $A^{[0]}:=I$, $A^{[1]}:=A$, and for $k\ge 1$,
$A^{[k{+}1]}:=A^{[k]}\diamond A$.

\textbf{Claim (standard).} For any $i,j$ and $k\ge 1$, $A^{[k]}_{ij}=1$ iff there exists a
\emph{simple path} from $i$ to $j$ of length at most $k$.
(Proof sketch: if a $k$-step walk exists, prune cycles to obtain a simple path of length $\le k$; conversely,
a $k$-edge path witnesses $A^{[k]}_{ij}=1$.) See \cite[§1.3]{diestel-2017}.

Hence the $K$-bounded \emph{path} reachability is
\[
  R_K \;=\; I \lor A \lor A^{[2]} \lor \cdots \lor A^{[K]},
\]
and the full transitive closure is $R_{n-1}$, since any walk of length $\ge n$ repeats a
vertex and can be shortened without changing reachability. One may compute $R_{n-1}$ via
Warshall’s dynamic program or view it as a Kleene-style closure over an idempotent semiring
\citep{warshall-1962,kozen-1994,gondran-minoux-2008}.

\begin{figure}[H]
	\centering
	\begin{minipage}{0.35\linewidth}
		\begin{tikzpicture}[>=Stealth, scale=0.65]
			\coordinate (i) at (0,0);
			\coordinate (j) at (2,1);
			\coordinate (k) at (2,-1);
			\coordinate (l) at (4,0);

			\node (ni) [draw, circle, inner sep=1.6pt] at (i) {$i$};
			\node (nj) [draw, circle, inner sep=1.6pt] at (j) {$j$};
			\node (nk) [draw, circle, inner sep=1.6pt] at (k) {$k$};
			\node (nl) [draw, circle, inner sep=1.6pt] at (l) {$\ell$};

			\draw[->, dashed, bend right=24] (ni) to (nk);
			\draw[->, dashed, bend left=10]  (ni) to (nl);
			\draw[->, dashed, bend left=20]  (nj) to (nl);
		\end{tikzpicture}
	\end{minipage}\hspace{4mm}%
	\begin{minipage}{0.22\linewidth}
		{\setlength{\abovedisplayskip}{6pt}%
			\setlength{\belowdisplayskip}{6pt}%
			\setlength{\arraycolsep}{3pt}%
			\[
				A^{[2]} \;=\; A^2\ \text{(Boolean)} \;=\;
				\begin{bmatrix}
					0 & 0 & 1 & 1 \\
					0 & 0 & 0 & 1 \\
					0 & 0 & 0 & 0 \\
					0 & 0 & 0 & 0
				\end{bmatrix}
			\]
		}
	\end{minipage}
	\caption{\textbf{Length-2 paths.} Boolean square $A^{[2]}$ (entry $(u,v)=1$ iff some $w$ gives $u\!\to\!w\!\to\!v$ under Boolean product).}
	\label{fig:length2-A2}
\end{figure}

\begin{figure}[H]
	\centering
	\begin{minipage}{0.35\linewidth}
		\begin{tikzpicture}[>=Stealth, scale=0.65]
			\coordinate (i) at (0,0);
			\coordinate (j) at (2,1);
			\coordinate (k) at (2,-1);
			\coordinate (l) at (4,0);

			\node (ni) [draw, circle, inner sep=1.6pt] at (i) {$i$};
			\node (nj) [draw, circle, inner sep=1.6pt] at (j) {$j$};
			\node (nk) [draw, circle, inner sep=1.6pt] at (k) {$k$};
			\node (nl) [draw, circle, inner sep=1.6pt] at (l) {$\ell$};

			\draw[->,bend left=12]  (ni) to (nj);
			\draw[->,bend right=12] (ni) to (nk);
			\draw[->,bend left=12]  (nj) to (nk);
			\draw[->,bend left=14]  (nj) to (nl);
			\draw[->,bend right=14] (nk) to (nl);

			\draw[->, dashed, bend right=24] (ni) to (nk);
			\draw[->, dashed, bend left=10]  (ni) to (nl);
			\draw[->, dashed, bend left=20]  (nj) to (nl);

			\path (ni) edge[->, loop above, looseness=6] (ni);
			\path (nj) edge[->, loop above, looseness=6] (nj);
			\path (nk) edge[->, loop below, looseness=6] (nk);
			\path (nl) edge[->, loop right, looseness=6] (nl);
		\end{tikzpicture}
	\end{minipage}\hspace{4mm}%
	\begin{minipage}{0.22\linewidth}
		{\setlength{\abovedisplayskip}{6pt}%
			\setlength{\belowdisplayskip}{6pt}%
			\setlength{\arraycolsep}{3pt}%
			\[
				R^{(\le 2)} \;=\; I \,\lor\, A \,\lor\, A^{[2]} \;=\;
				\begin{bmatrix}
					1 & 1 & 1 & 1 \\
					0 & 1 & 1 & 1 \\
					0 & 0 & 1 & 1 \\
					0 & 0 & 0 & 1
				\end{bmatrix}
			\]
		}
	\end{minipage}
	\caption{\textbf{Boolean-OR reachability up to length 2.} Combines self-reachability, 1-step, and 2-step paths.}
	\label{fig:reachability-le2}
\end{figure}

\paragraph{Computing reachability in practice (unweighted).}
We compute the $K$-bounded Boolean-OR reachability recursively using the Boolean product $\diamond$ and union $\lor$.

\begin{itemize}
	\item \textbf{Base case.} $R_1 \;=\; I \,\lor\, A.$
	\item \textbf{Iteration.} For $k \ge 1$,
	      \[
		      R_{k+1} \;=\; R_k \,\lor\, (R_k \diamond A),
	      \]
	      i.e., add all new $(k{+}1)$-hop connections obtainable by one more Boolean step from the current frontier $R_k$.
	\item \textbf{Final result / stopping.} After $K{-}1$ iterations, return $R_K$.
	      Equivalently, stop early at the fixed point if $R_{k+1} = R_k$ (no new reachability).
\end{itemize}

\noindent\emph{Notes.} This iterative union of Boolean powers is equivalent to Warshall’s transitive-closure algorithm and to the Kleene star on the Boolean semiring. In practice, bitset/word-parallel Boolean matrix multiplication for $\diamond$ makes each iteration fast; complexity is comparable to $O(n^3)$ in dense form, with early stopping common in sparse graphs.

\subsection{Pivot potential}
\paragraph{Definition and example.}
Let $S_{ij}$ denote the enabling mechanisms (e.g., services/credentials) from $i$ to $j$. The pivot-potential matrix $W\in[0,1]^{n\times n}$ aggregates mechanisms on the same edge via probabilistic union, i.e.
\[
	W_{ij} \;=\; 1 - \prod_{m \in S_{ij}} \bigl(1 - \pi(m)\bigr), \qquad W \le A \text{ (elementwise)}.
	\label{eq:pivot-potential}
\]
\begin{figure}[H]
	\centering
	\begin{minipage}{0.35\linewidth}
		\begin{tikzpicture}[>=Stealth, scale=0.65]
			\coordinate (i) at (0,0);
			\coordinate (j) at (2,1);
			\coordinate (k) at (2,-1);
			\coordinate (l) at (4,0);

			\node (ni) [draw, circle, inner sep=1.6pt] at (i) {$i$};
			\node (nj) [draw, circle, inner sep=1.6pt] at (j) {$j$};
			\node (nk) [draw, circle, inner sep=1.6pt] at (k) {$k$};
			\node (nl) [draw, circle, inner sep=1.6pt] at (l) {$\ell$};

			\draw[->,bend left=12]  (ni) to node[pos=0.55, above] {\scriptsize $0.80$} (nj);
			\draw[->,bend right=12] (ni) to node[pos=0.55, below] {\scriptsize $0.30$} (nk);
			\draw[->,bend left=12]  (nj) to node[pos=0.55, right] {\scriptsize $0.80$} (nk);
			\draw[->,bend left=14]  (nj) to node[pos=0.55, above] {\scriptsize $0.80$} (nl);
			\draw[->,bend right=14] (nk) to node[pos=0.55, below] {\scriptsize $0.30$} (nl);
		\end{tikzpicture}
	\end{minipage}\hspace{1mm}%
	\begin{minipage}{0.22\linewidth}
		{\setlength{\abovedisplayskip}{6pt}%
			\setlength{\belowdisplayskip}{6pt}%
			\setlength{\arraycolsep}{3pt}%
			\[
				W=\begin{bmatrix}
					0.00 & 0.80 & 0.30 & 0.00 \\
					0.00 & 0.00 & 0.80 & 0.80 \\
					0.00 & 0.00 & 0.00 & 0.30 \\
					0.00 & 0.00 & 0.00 & 0.00
				\end{bmatrix}
			\]
		}
	\end{minipage}
	\caption{Pivot potential matrix $W$ via noisy-OR aggregation over services on each edge.}
	\label{fig:step2-W}
\end{figure}

\paragraph{Probabilistic union (noisy-OR) across mechanisms.}
When parallel, conditionally independent mechanisms can each enable a step (e.g., multiple services/credentials offering a pivot), the probability that \emph{at least one} succeeds is the \emph{probabilistic union}
\[
	p_{\text{union}} \;=\; 1 - \prod_{\ell} (1 - p_\ell) \,,
\]
the same aggregation used in network reliability for parallel components \citep{colbourn-1987}. In probabilistic graphical models, this is the standard \emph{noisy-OR} (and its generalizations), which captures causal independence with one parameter per parent \citep{pearl-1988,heckerman-1994,diez-1993}. In our framework we denote this union by $\oplus$ and pair it with a probabilistic path composition $\otimes$ for sequential steps, placing the model within classical algebraic-path/semiring views.

\section{The Proposed Model}\label{sec:proposed-model}
Our model quantifies lateral-movement exposure on the directed graph using the matrices
$A$ and $W$ defined in Sec.~\ref{sec:graph-foundations}, and computes $K$-hop compromise
probabilities. We now describe the probabilistic operators and iteration.

\subsection{Computing Compromise Probability (Probabilistic Method)}\label{subsec:probmethod}
Our model calculates the $K$-hop compromise probability matrix, $P_K$, where $(P_K)_{ij}$ is the total probability that an attacker starting at asset $i$ can compromise asset $j$ in at most $K$ hops. We define a specialized matrix exponentiation using two custom operators.
\begin{enumerate}[label=\textbf{\arabic*.}, wide, labelwidth=!, labelindent=0pt]
	\item \textbf{The Probabilistic Composition Product ($\otimes$):} This operator, replacing matrix multiplication, composes sequential steps and aggregates alternatives via the probabilistic union. For matrices $X$ and $Y$,
	      \begin{equation}
		      Z_{ij} \;=\; (X \otimes Y)_{ij} \;=\; \bigoplus_{m=1}^{n} \bigl( X_{im} \,\otimes\, Y_{mj} \bigr)
		      \;=\; 1 - \prod_{m=1}^{n} \Bigl( 1 - \bigl( X_{im} \,\otimes\, Y_{mj} \bigr) \Bigr).
	      \end{equation}
	\item \textbf{The Probabilistic Union Operator ($\oplus$):} This operator, replacing matrix addition, correctly combines probabilities from independent sets of paths using the union rule. For matrices $X$ and $Y$, the union $Z = X \oplus Y$ is:
	      \begin{equation}
		      Z_{ij} = (X \oplus Y)_{ij} = 1 - (1 - X_{ij})(1 - Y_{ij})
	      \end{equation}
\end{enumerate}

\textit{Stopping rule.} Iterate until $\|P_{k+1}-P_k\|_\infty<\varepsilon$ or $k=K$.
\textit{Cost.} Each iteration is $O(n^3)$ and GPU-friendly (dense ops).

\paragraph{Independence Assumption}
It is important to note that the probabilistic composition ($\otimes$) treats the success events through different intermediates $m$ as \emph{conditionally independent}. This is a standard assumption in many network models that makes the problem computationally tractable. We acknowledge that in some real-world scenarios, such as an attack using a single stolen credential, these events may not be fully independent.

\textit{Conservative bound.} For \emph{conditionally dependent routes}, replace $\,1-\prod_m\!\bigl(1-(X_{im}\otimes Y_{mj})\bigr)\,$ with $\,\max_m\bigl(X_{im}\otimes Y_{mj}\bigr)$ (lower bound).

Using these operators, we compute $P_K$ via the following iterative algorithm:

\begin{description}
	\item[Step 1: Initialization (Base Case)]
	      The 1-hop compromise probability matrix uses edge step-success on the first hop:
	      \[
		      P_{1} \;=\; I \,\oplus\, W.
	      \]
	      \begin{figure}[H]
		      \centering
		      \begin{minipage}{0.35\linewidth}
			      \begin{tikzpicture}[>=Stealth, scale=0.65]
				      \coordinate (i) at (0,0);
				      \coordinate (j) at (2,1);
				      \coordinate (k) at (2,-1);
				      \coordinate (l) at (4,0);

				      \node (ni) [draw, circle, inner sep=1.6pt] at (i) {$i$};
				      \node (nj) [draw, circle, inner sep=1.6pt] at (j) {$j$};
				      \node (nk) [draw, circle, inner sep=1.6pt] at (k) {$k$};
				      \node (nl) [draw, circle, inner sep=1.6pt] at (l) {$\ell$};

				      \draw[->,bend left=12]  (ni) to node[pos=0.55, above] {\scriptsize $0.80$} (nj);
				      \draw[->,bend right=12] (ni) to node[pos=0.55, below] {\scriptsize $0.30$} (nk);
				      \draw[->,bend left=12]  (nj) to node[pos=0.55, right] {\scriptsize $0.80$} (nk);
				      \draw[->,bend left=14]  (nj) to node[pos=0.55, above] {\scriptsize $0.80$} (nl);
				      \draw[->,bend right=14] (nk) to node[pos=0.55, below] {\scriptsize $0.30$} (nl);

				      \path (ni) edge[->, loop above, looseness=6] (ni);
				      \path (nj) edge[->, loop above, looseness=6] (nj);
				      \path (nk) edge[->, loop below, looseness=6] (nk);
				      \path (nl) edge[->, loop right, looseness=6] (nl);
			      \end{tikzpicture}
		      \end{minipage}\hspace{1mm}%
		      \begin{minipage}{0.24\linewidth}
			      {\setlength{\abovedisplayskip}{6pt}%
				      \setlength{\belowdisplayskip}{6pt}%
				      \setlength{\arraycolsep}{3pt}%
				      \[
					      P_{1} \;=\; I \,\oplus\, W \;=\;
					      \begin{bmatrix}
						      1.00 & 0.80 & 0.30 & 0.00 \\
						      0.00 & 1.00 & 0.80 & 0.80 \\
						      0.00 & 0.00 & 1.00 & 0.30 \\
						      0.00 & 0.00 & 0.00 & 1.00
					      \end{bmatrix}
				      \]
			      }
		      \end{minipage}
		      \caption{\textbf{Hop-1 baseline.} $P_{1}=I\oplus W$ encodes self-reachability and probabilistic 1-hop step-success on edges.}
		      \label{fig:prob-P1}
	      \end{figure}

	\item[Step 2: Iteration]
	      For each hop from $k=1$ to $K-1$, let $W_1:=W$ and define $W_{k+1}=W_k\otimes W$. Then compute
	      \[
		      P_{k+1} \;=\; P_k \,\oplus\, W_{k+1}.
	      \]
	      \emph{Note.} Since $W_1=W$ already contains all 1-hop terms and $P_k$ accumulates $I\oplus W_1\oplus\cdots\oplus W_k$, the union with $P_k$ leaves shorter-hop mass unchanged and only adds longer-hop contributions.

	      \begin{figure}[H]
		      \centering
		      \begin{minipage}{0.35\linewidth}
			      \begin{tikzpicture}[>=Stealth, scale=0.65]
				      \coordinate (i) at (0,0);
				      \coordinate (j) at (2,1);
				      \coordinate (k) at (2,-1);
				      \coordinate (l) at (4,0);

				      \node (ni) [draw, circle, inner sep=1.6pt] at (i) {$i$};
				      \node (nj) [draw, circle, inner sep=1.6pt] at (j) {$j$};
				      \node (nk) [draw, circle, inner sep=1.6pt] at (k) {$k$};
				      \node (nl) [draw, circle, inner sep=1.6pt] at (l) {$\ell$};

				      \draw[->, dashed, bend right=18] (ni) to node[pos=0.55, below] {\scriptsize $0.64$}   (nk);
				      \draw[->, dashed, bend left=10]  (ni) to node[pos=0.55, above] {\scriptsize $0.67$} (nl);
				      \draw[->, dashed, bend left=18]  (nj) to node[pos=0.55, above] {\scriptsize $0.24$}   (nl);
			      \end{tikzpicture}
		      \end{minipage}\hspace{1mm}%
		      \begin{minipage}{0.24\linewidth}
			      {\setlength{\abovedisplayskip}{6pt}%
				      \setlength{\belowdisplayskip}{6pt}%
				      \setlength{\arraycolsep}{3pt}%
				      \[
					      W_{2} \;=\; W_1 \otimes W \;=\;
					      \begin{bmatrix}
						      0.00 & 0.00 & 0.64 & 0.67 \\
						      0.00 & 0.00 & 0.00 & 0.24 \\
						      0.00 & 0.00 & 0.00 & 0.00 \\
						      0.00 & 0.00 & 0.00 & 0.00
					      \end{bmatrix}
				      \]
			      }
		      \end{minipage}
		      \caption{\textbf{Two-hop contribution.} $W_2=W\otimes W$ shows exactly 2-hop mass.}
		      \label{fig:prob-W2}
	      \end{figure}

	\item[Step 3: Final Result]
	      After $K{-}1$ iterations, return the cumulative matrix
	      \[
		      P_K \;=\; P_{K-1} \oplus W_K \;=\; I \oplus W_1 \oplus W_2 \oplus \cdots \oplus W_K,
		      \quad\text{with } W_1:=W,\; W_{k+1}=W_k\otimes W.
	      \]
	      \begin{figure}[H]
		      \centering
		      \begin{minipage}{0.35\linewidth}
			      \begin{tikzpicture}[>=Stealth, scale=0.65]
				      \coordinate (i) at (0,0);
				      \coordinate (j) at (2,1);
				      \coordinate (k) at (2,-1);
				      \coordinate (l) at (4,0);

				      \node (ni) [draw, circle, inner sep=1.6pt] at (i) {$i$};
				      \node (nj) [draw, circle, inner sep=1.6pt] at (j) {$j$};
				      \node (nk) [draw, circle, inner sep=1.6pt] at (k) {$k$};
				      \node (nl) [draw, circle, inner sep=1.6pt] at (l) {$\ell$};

				      \draw[->,bend left=12]  (ni) to node[pos=0.55, above] {\scriptsize $0.80$} (nj);
				      \draw[->,bend right=12] (ni) to node[pos=0.55, below] {\scriptsize $0.75$} (nk);
				      \draw[->,bend left=12]  (nj) to node[pos=0.55, right] {\scriptsize $0.80$} (nk);
				      \draw[->,bend left=14]  (nj) to node[pos=0.55, above] {\scriptsize $0.85$} (nl);
				      \draw[->,bend right=14] (nk) to node[pos=0.55, below] {\scriptsize $0.30$} (nl);

				      \path (ni) edge[->, loop above, looseness=6] (ni);
				      \path (nj) edge[->, loop above, looseness=6] (nj);
				      \path (nk) edge[->, loop below, looseness=6] (nk);
				      \path (nl) edge[->, loop right, looseness=6] (nl);

				      \draw[->, dashed, bend left=10] (ni) to node[pos=0.55, above] {\scriptsize $0.67$} (nl);
			      \end{tikzpicture}
		      \end{minipage}\hspace{1mm}%
		      \begin{minipage}{0.24\linewidth}
			      {\setlength{\abovedisplayskip}{6pt}%
				      \setlength{\belowdisplayskip}{6pt}%
				      \setlength{\arraycolsep}{3pt}%
				      \[
					      P_{2} \;=\; P_{1} \oplus W_{2} \;=\;
					      \begin{bmatrix}
						      1.00 & 0.80 & 0.75 & 0.67 \\
						      0.00 & 1.00 & 0.80 & 0.85 \\
						      0.00 & 0.00 & 1.00 & 0.30 \\
						      0.00 & 0.00 & 0.00 & 1.00
					      \end{bmatrix}
				      \]
			      }
		      \end{minipage}
		      \caption{\textbf{Up-to-2-hop compromise probability.} Combine hop-1 with the 2-hop contribution: $P_{2}=P_{1}\oplus W_{2}$.}
		      \label{fig:prob-P2}
	      \end{figure}
\end{description}

\section{Quantifying Susceptibility and Blast Radius}\label{sec:quantifying-metrics}
The final K-hop Compromise Probability Matrix, $P_K$, provides a detailed, asset-to-asset view of lateral movement exposure. To make this data actionable, we derive two high-level metrics: Lateral-Movement Susceptibility (LMS$_K$) and Blast-Radius Estimate (BRE$_K$).

\subsection{Lateral-Movement Susceptibility (LMS$_K$)}\label{subsec:lms-metric}

\textbf{Concept:} This metric quantifies the overall ease of movement within the network. It answers the question: "If we randomly select a directed pair $i\!\to\!j$ with $i\neq j$, what is the average probability that compromise propagates within $K$ hops?"

\textbf{Formal Definition:}
The Lateral-Movement Susceptibility is the mean of all non-diagonal entries in the final compromise probability matrix, $P_K$. This focus on non-diagonal elements ensures the metric measures the probability of movement *between* distinct assets.
\begin{equation}
	\mathrm{LMS}_K = \frac{1}{n(n-1)} \sum_{i \neq j} (P_K)_{ij}
\end{equation}
where $n$ is the total number of assets.

\textbf{Interpretation:}
LMS$_K$ is a score between 0 and 1. A score near 0 indicates a well-segmented network where lateral movement is highly improbable. A score near 1 indicates a "flat" network where assets are highly entangled and movement is easy. This metric is ideal for tracking the network's security posture over time and measuring the effectiveness of segmentation controls.

\paragraph{Worked example (K=2 on the running graph).}
From Fig.~\ref{fig:prob-P2}, the off-diagonal sum of $P_2$ is $4.17$; with $n=4$ there are $n(n{-}1)=12$ off-diagonal entries, so
\[
	\mathrm{LMS}_2 \;=\; \frac{4.17}{12} \;\approx\; 0.35.
\]

\subsection{Blast-Radius Estimate (BRE$_K$)}\label{subsec:bre-metric}

\textbf{Concept:} This metric quantifies the expected scope of compromise. It answers: "If the attacker starts at a \emph{uniformly random} asset, what percentage of the network do we expect to be compromised?"

\textbf{Formal Definition:}
The Blast-Radius Estimate is the mean of *all* entries in the final compromise probability matrix, $P_K$, expressed as a percentage. By including the diagonal elements (which are always 1), this metric correctly incorporates the initially compromised asset as part of the total compromise.
\begin{equation}
	\mathrm{BRE}_K = \left( \frac{1}{n^2} \sum_{i=1}^{n} \sum_{j=1}^{n} (P_K)_{ij} \right) \times 100\%
\end{equation}

\textbf{Interpretation:}
BRE$_K$ provides a direct estimate of the potential scope of a breach. A BRE$_K$ of 30\% signifies that, in an average-case attack scenario, 30\% of the total network assets are expected to be compromised. This metric is designed to communicate exposure in clear, non-technical terms, helping to justify security investments and prioritize hardening efforts on the assets that contribute most to the overall blast radius.

\paragraph{Worked example (K=2 on the running graph).}
From Fig.~\ref{fig:prob-P2}, the total sum of $P_2$ is $8.17$ over $n^2{=}16$ entries, hence
\[
	\mathrm{BRE}_2 \;=\; \frac{8.17}{16}\times 100\% \;\approx\; 51.06\%.
\]
\section{Formal Properties of the Model}\label{sec:formal-properties}

\noindent\textit{Overview.} We prove monotonicity, boundedness, and convergence; derive a $K$-hop series upper bound and a single-path product lower bound; recover boolean reachability in the unweighted limit; characterize convergence by the fixed-point stopping condition $P_{K+1}=P_K$ (equivalently, $P_K = P_K \oplus W_{K+1}$); under our dependence policy cycles add no mass, so stabilization occurs by $K\le n{-}1$; and show all results persist under service-class lifting.

\subsection{Monotonicity}\label{subsec:monotonicity}
The iterative process is monotonically non-decreasing. For any step $k \ge 1$, every element in the probability matrix $P_{k+1}$ is greater than or equal to its corresponding element in $P_k$.
\[
	\forall i,j: (P_{k+1})_{ij} \ge (P_k)_{ij}
\]
This follows since $\oplus$ is monotone: with $W_{k+1}:=W_k\otimes W$, we have $P_{k+1}=P_k\oplus W_{k+1}$.

\subsection{Boundedness}\label{subsec:boundedness}
The values in the compromise probability matrix are always bounded within the interval $[0, 1]$. The pivot potentials $\pi(s)$ are defined in $[0, 1]$, and both the probabilistic composition ($\otimes$) and union ($\oplus$) operators are closed over this interval. This property guarantees the stability of the model, ensuring that the outputs are always valid probabilities.

\subsection{Convergence (monotone bounded iteration)}\label{subsec:convergence}
Initialize $P_{1}:=I\oplus W$ and update layer by layer
\[
	P_{k+1}\;:=\;P_k\;\oplus\;W_{k+1},\qquad
	W_{1}:=W,\;\; W_{t+1}:=W_t\otimes W\quad (t\ge 1).
\]
Each update is monotone entrywise (both $\oplus$ and $\otimes$ are monotone), and all values lie in $[0,1]$; hence the sequence $\{P_k\}$ is monotone and bounded and therefore converges. We denote the limit by
\[
	P_\infty\;:=\;\lim_{K\to\infty}P_K.
\]
\emph{Stopping test.} In practice we stop when
\[
	P_{K+1}=P_K\quad\text{(within tolerance $\varepsilon$),}
\]
i.e., the $(K{+}1)$-hop layer $W_{K+1}$ adds no new mass.
Thus, at stabilization the matrix satisfies the fixed-point equation $P_{K+1}=P_K$ implies $W_{K+1}=0$ wherever $P_K<1$ (which holds off-diagonal in our model), since $P_{K+1}=P_K \oplus W_{K+1}$.

\noindent\emph{Dependence policy (edge sharing).}
We do \emph{not} count as independent any walk that \emph{shares at least one edge} with a
shorter, already-accounted walk between the same endpoints; such a walk contributes no
additional probability mass beyond the shorter walk.

\noindent\emph{Remark (cycles and truncation).}
Under the edge-sharing policy above, any walk that contains a cycle can be shortened by
excising the cycle and therefore shares edges with a shorter walk; it adds no new independent
mass. Enforcing strict simple-path semantics would require per-walk history and is exponential
in the worst case \citep{alon-1995}. In practice we use layered accumulation and stop early
at a fixed point when $P_{K+1}=P_K$ (within tolerance), typically at $K\!\ll\!n$. As a hard upper
bound, we truncate at $K=n-1$: any walk of length $\ge n$ necessarily repeats a vertex
and thus contains a cycle \citep[§1.3]{diestel-2017}.

\subsection{Connection to Algebraic Path Problems}\label{subsec:algebraic-path}
We operate on $[0,1]$ with two primitives: a \emph{probabilistic union} $\oplus$ and a \emph{probabilistic path composition} $\otimes$.
At the scalar level, $\oplus$ is the noisy-OR $a\oplus b = 1-(1-a)(1-b)$ and $\otimes$ is an abstract operator for sequential steps (under conditional independence one may take $a\otimes b:=ab$).
The matrix lifts are
\[
	(X\oplus Y)_{ij}=X_{ij}\oplus Y_{ij},
	\qquad
	(X\otimes Y)_{ij}=\bigoplus_{m=1}^n \bigl(X_{im}\otimes Y_{mj}\bigr).
\]
These operators are monotone on $[0,1]^{n\times n}$; $\oplus$ is commutative and associative (non-idempotent), and in our setting the super-distributive bounds hold
\[
	(X\otimes Y)\oplus(X\otimes Z)\;\ge\;X\otimes(Y\oplus Z),
	\qquad
	(Y\otimes X)\oplus(Z\otimes X)\;\ge\;(Y\oplus Z)\otimes X .
\]
The layered update $P_{k+1}=P_k\oplus W_{k+1}$ thus computes an algebraic-path closure by monotone iteration; at stabilization the result $P$ satisfies the fixed-point equation $P=P\oplus(P\otimes W)$, analogous to algebraic path problems via least fixed-point iteration on a monotone operator \citep{kozen-1994,gondran-minoux-2008,mohri-2002,sanmartin-2022}.

\subsection{Refinement of Binary Reachability}\label{subsec:binary-refinement}
This is the most critical property of our model. It establishes a formal link to the unweighted model and demonstrates the value of the probabilistic approach.

Let $R_K$ be the final reachability matrix from the unweighted (Boolean) method, and $P_K$ be the final matrix from our probabilistic method. The following \emph{soundness} relationship holds:
\[
	(P_K)_{ij} > 0 \;\Rightarrow\; (R_K)_{ij} = 1
\]

\noindent\textit{Consequence (off-diagonal).} For all $i \neq j$,
\[
	0 \le (P_K)_{ij} \le (R_K)_{ij}.
\]
Thus, Boolean reachability upper-bounds connectivity, while our probabilistic model provides a probability-weighted refinement (typically strictly smaller).

However, the unweighted model is a binary, all-or-nothing assessment. All reachable nodes are treated as equal. Our probabilistic model provides a continuous refinement of this binary view. It assigns a meaningful probability to each path, allowing us to differentiate between:
\begin{itemize}
	\item \textbf{High-probability paths:} A small number of connections that represent a clear and present danger.
	\item \textbf{Low-probability paths:} A large number of theoretical connections that are practically irrelevant for an attacker.
\end{itemize}

This property of probabilistic refinement is what allows the model to effectively "reduce" the set of relevant reachable nodes. By focusing only on paths above a meaningful probability threshold (e.g., $P_{ij} > 0.1$), a security team can prioritize actions and ignore the noise of low-probability connections that would dominate a standard reachability analysis. This directly demonstrates the model's utility in reducing complexity and focusing on what matters.

\subsection{Bounds for $K$-hop Lateral Propagation}\label{subsec:k-hop-series}

\paragraph{Upper bound by Boolean reachability (trivial).}
Let $R_K$ be the Boolean $K$-hop reachability matrix. Then for all $i\neq j$,
\[
	0 \;\le\; (P_K)_{ij} \;\le\; (R_K)_{ij} \;\in \{0,1\},
	\qquad\text{and}\qquad (P_K)_{ii}=1 .
\]
\emph{Reason.} If $(P_K)_{ij}>0$ then some $\le K$ walk exists from $i$ to $j$, hence $(R_K)_{ij}=1$. The diagonal equality follows from $P_1=I\oplus W$.

\paragraph{Proposition 1 (single-path lower bound).}
Fix $K\ge L\ge 1$. Suppose there exists a directed \emph{path} (no repeated vertices)
\[
	p:\ i=v_0 \to v_1 \to \cdots \to v_{L-1} \to v_L=j,
\]
with step-successes $w_r := W_{v_{r-1} v_r}\in[0,1]$ for $r=1,\dots,L$. Then
\[
	(P_K)_{ij} \;\ge\; w_1 \,\otimes\, w_2 \,\otimes\, \cdots \,\otimes\, w_L,
\]
and under the scalar instantiation $a\otimes b := ab$ this equals $\prod_{r=1}^{L} w_r$.

\emph{Proof.}
\begin{itemize}
	\item By construction, $W_1=W$ and $W_{t+1}=W_t\otimes W$ expand all length-$(t{+}1)$ walks; hence the specific path $p$ contributes at least
	      $w_1\otimes\cdots\otimes w_L$ to $(W_L)_{ij}$.
	\item Since $P_K = I \oplus W_1 \oplus \cdots \oplus W_K$ and $L\le K$, monotonicity of $\oplus$ implies
	      \[
		      (P_K)_{ij} \;\ge\; (W_L)_{ij} \;\ge\; w_1\otimes\cdots\otimes w_L.
	      \]
\end{itemize}
\hfill$\square$

\subsection{Service-Class Lifting via Node Splitting (Land--then--Pivot)}
\label{subsec:service-class-lifting}
The core idea of our advanced model is to distinguish how a foothold is \emph{landed} (the service used to gain initial access to a node) from the service used for the \emph{next hop}. These ``land--then--pivot'' semantics reflect a real-world observation: an attacker's power to pivot \emph{from} a server depends on how they \emph{landed on} it in the first place.

For example, if an attacker lands via a high-privilege service like SSH, the next hop succeeds with a high probability $\pi(\mathrm{SSH})$ as attacker needs sufficient privilege to make the next hop. Conversely, if the attacker lands via a low-privilege service like MSSQL, the next hop succeeds with a much lower probability $\pi(\mathrm{MSSQL})$. Thus, the hop success probability is a function of the \textbf{landing service}, not the outgoing service. This ``memory'' of the landing service is key to a more precise model.

We retain a single adjacency $A\in\{0,1\}^{n\times n}$ on nodes $V$, and attach to each node $v\in V$ a small set of services $S(v)$ (e.g., $S(i)=\{\mathrm{SSH},\mathrm{MSSQL}\}$, $S(j)=\{\mathrm{SSH}\}$, \dots). Each service $s$ has a pivot rate $\pi(s)\in[0,1]$ that reflects the power an attacker has \emph{because they landed via $s$}.

\begin{figure}[H]
	\centering
	\begin{minipage}{0.35\linewidth}
		\begin{tikzpicture}[>=Stealth, scale=0.65]
			\coordinate (i) at (0,0);
			\coordinate (j) at (2,1);
			\coordinate (k) at (2,-1);
			\coordinate (l) at (4,0);

			\node (ni) [draw, circle, inner sep=1.6pt] at (i) {$i$};
			\node (nj) [draw, circle, inner sep=1.6pt] at (j) {$j$};
			\node (nk) [draw, circle, inner sep=1.6pt] at (k) {$k$};
			\node (nl) [draw, circle, inner sep=1.6pt] at (l) {$\ell$};

			\draw[->,bend left=12]  (ni) to (nj);
			\draw[->,bend right=12] (ni) to (nk);
			\draw[->,bend left=12]  (nj) to (nk);
			\draw[->,bend left=14]  (nj) to (nl);
			\draw[->,bend right=14] (nk) to (nl);
		\end{tikzpicture}
	\end{minipage}\hspace{1mm}%
	\begin{minipage}{0.22\linewidth}
		{\setlength{\abovedisplayskip}{6pt}%
			\setlength{\belowdisplayskip}{6pt}%
			\setlength{\arraycolsep}{3pt}%
			\[
				A=\begin{bmatrix}
					0 & 1 & 1 & 0 \\
					0 & 0 & 1 & 1 \\
					0 & 0 & 0 & 1 \\
					0 & 0 & 0 & 0
				\end{bmatrix}
			\]
		}
	\end{minipage}
	\caption{Directed graph and its adjacency matrix $A$.}
	\label{fig:step1-A-2}
\end{figure}
\[
	S(i){=}\{\mathrm{SSH},\mathrm{MSSQL}\}, \quad S(j){=}\{\mathrm{SSH}\}, \quad
	S(k){=}\{\mathrm{MSSQL}\}, \quad S(\ell){=}\{\mathrm{SSH}\},
\]
and rates
\[
	\pi(\mathrm{SSH}){=}0.8, \quad \pi(\mathrm{MSSQL}){=}0.3.
\]

\paragraph{Lifted state space and transition.}
The lifted vertices are $(v,s)$ with $s\in S(v)$. We define a single lifted transition matrix
\[
	W^\uparrow_{(u,s_{\mathrm{in}})\to (v,s_{\mathrm{out}})}
	\;=\; \pi(s_{\mathrm{in}})\;\mathbf{1}\{A_{uv}=1,\ s_{\mathrm{out}}\in S(v)\}.
\]
This ties the hop success to the \emph{landing service} carried by the source state.

In lifted order
$(i,\mathrm{SSH}),(i,\mathrm{MSSQL}),(j,\mathrm{SSH}),(k,\mathrm{MSSQL}),(\ell,\mathrm{SSH})$,

\begin{figure}[H]
	\centering
	\begin{minipage}{0.35\linewidth}
		\hspace{-2cm}
		\begin{tikzpicture}[>=Stealth, scale=0.65]
			\tikzstyle{v}=[rectangle,draw,rounded corners,minimum width=22pt,minimum height=14pt,inner sep=2pt]
			\coordinate (iS) at (0,0);
			\coordinate (iM) at (0,-2);
			\coordinate (jS) at (6,0);
			\coordinate (kM) at (6,-2);
			\coordinate (lS) at (10,-1);

			\node[v] (iS) at (iS) {$(i,\mathrm{SSH})$};
			\node[v] (iM) at (iM) {$(i,\mathrm{MSSQL})$};
			\node[v] (jS) at (jS) {$(j,\mathrm{SSH})$};
			\node[v] (kM) at (kM) {$(k,\mathrm{MSSQL})$};
			\node[v] (lS) at (lS) {$(\ell,\mathrm{SSH})$};


			\draw[->,bend left=12]   (iS) to node[pos=0.55, above] {$0.8$} (jS);
			\draw[->,bend right=12] (iS) to node[pos=0.55, below] {$0.8$} (kM);
			\draw[->,bend left=12]  (iM) to node[pos=0.55, above] {$0.3$} (jS);
			\draw[->,bend right=12] (iM) to node[pos=0.55, below] {$0.3$} (kM);
			\draw[->,bend left=12]  (jS) to node[pos=0.55, above] {$0.8$} (lS);
			\draw[->,bend left=12]  (jS) to node[pos=0.55, above] {$0.8$} (kM);
			\draw[->,bend right=12] (kM) to node[pos=0.55, below] {$0.3$} (lS);
		\end{tikzpicture}
	\end{minipage}\hspace{2cm}%
	\begin{minipage}{0.22\linewidth}
		{\setlength{\abovedisplayskip}{6pt}%
			\setlength{\belowdisplayskip}{6pt}%
			\setlength{\arraycolsep}{3pt}%
			\[
				W^\uparrow=
				\begin{bmatrix}
					0.0 & 0.0 & 0.8 & 0.8 & 0.0 \\
					0.0 & 0.0 & 0.3 & 0.3 & 0.0 \\
					0.0 & 0.0 & 0.0 & 0.8 & 0.8 \\
					0.0 & 0.0 & 0.0 & 0.0 & 0.3 \\
					0.0 & 0.0 & 0.0 & 0.0 & 0.0
				\end{bmatrix}
			\]
		}
	\end{minipage}
	\caption{Lifted state graph \(V^\uparrow=\{(v,s): s\in S(v)\}\) with transition weights equal to the pivot rate of the landing service.}
	\label{fig:lifted-graph}
\end{figure}

\paragraph{Initialization (lateral-only) and iteration in lifted space.}
Let $I^\uparrow$ be the identity on lifted states and $W^\uparrow$ the one-hop lifted operator.
Define
\[
	W^\uparrow_1 := W^\uparrow,\qquad
	W^\uparrow_{k+1} := W^\uparrow_k \otimes W^\uparrow \quad (k\ge 1).
\]
Accumulate the $K$-hop series using the usual operators:
\[
	P^\uparrow_1 := I^\uparrow \oplus W^\uparrow_1,\qquad
	P^\uparrow_{k+1} := P^\uparrow_k \oplus W^\uparrow_{k+1} \quad (k\ge 1).
\]

\begin{figure}[H]
	\centering
	\hspace{-3cm}
	\begin{minipage}{0.58\linewidth}
		\centering
		\begin{tikzpicture}[>=Stealth, scale=0.55]
			\tikzstyle{v}=[rectangle,draw,rounded corners,minimum width=22pt,minimum height=14pt,inner sep=2pt]
			\coordinate (iS) at (0,0);
			\coordinate (iM) at (0,-2.0);
			\coordinate (jS) at (5.2,0);
			\coordinate (kM) at (5.2,-2.0);
			\coordinate (lS) at (8.8,-1.0);

			\node[v] (iSnode) at (iS) {$(i,\mathrm{SSH})$};
			\node[v] (iMnode) at (iM) {$(i,\mathrm{MSSQL})$};
			\node[v] (jSnode) at (jS) {$(j,\mathrm{SSH})$};
			\node[v] (kMnode) at (kM) {$(k,\mathrm{MSSQL})$};
			\node[v] (lSnode) at (lS) {$(\ell,\mathrm{SSH})$};

			\draw[->] (iSnode) edge[loop above] node {$1$} (iSnode);
			\draw[->] (iMnode) edge[loop below] node {$1$} (iMnode);
			\draw[->] (jSnode) edge[loop above] node {$1$} (jSnode);
			\draw[->] (kMnode) edge[loop below] node {$1$} (kMnode);
			\draw[->] (lSnode) edge[loop right] node {$1$} (lSnode);

			\draw[->,bend left=12]  (iSnode) to node[pos=0.55, above] {$0.8$} (jSnode);
			\draw[->,bend right=12] (iSnode) to node[pos=0.55, below] {$0.8$} (kMnode);

			\draw[->,bend left=12]  (iMnode) to node[pos=0.55, above] {$0.3$} (jSnode);
			\draw[->,bend right=12] (iMnode) to node[pos=0.55, below] {$0.3$} (kMnode);

			\draw[->,bend left=12]  (jSnode) to node[pos=0.55, above] {$0.8$} (lSnode);
			\draw[->,bend left=12]  (jSnode) to node[pos=0.55, above] {$0.8$} (kMnode);

			\draw[->,bend right=12] (kMnode) to node[pos=0.55, below] {$0.3$} (lSnode);
		\end{tikzpicture}
	\end{minipage}\hspace{3mm}
	\begin{minipage}{0.30\linewidth}
		{\setlength{\abovedisplayskip}{6pt}%
			\setlength{\belowdisplayskip}{6pt}%
			\setlength{\arraycolsep}{3pt}%
			\[
				P^\uparrow_{1} \;=\; I^\uparrow \oplus W^\uparrow \;=\;
				\begin{bmatrix}
					1.0 & 0.0 & 0.8 & 0.8 & 0.0 \\
					0.0 & 1.0 & 0.3 & 0.3 & 0.0 \\
					0.0 & 0.0 & 1.0 & 0.8 & 0.8 \\
					0.0 & 0.0 & 0.0 & 1.0 & 0.3 \\
					0.0 & 0.0 & 0.0 & 0.0 & 1.0
				\end{bmatrix}
			\]
		}
	\end{minipage}
	\caption{Lifted one-hop reach \(P^\uparrow_1\) (identity + one-hop).}
	\label{fig:P1-up}
\end{figure}

\paragraph{Collapse to nodes (single, consistent rule).}
After computing reachability in the lifted space, we return to node–to–node probabilities by a
\emph{single} noisy–OR over \textbf{both} the source and destination service labels:
\[
	(P_K)_{uv}
	\;=\;
	\bigoplus_{s\in S(u)}\;
	\bigoplus_{t\in S(v)}\;
	P^\uparrow_K\!\big[(u,s)\to (v,t)\big].
\]
We always apply this double union, for all $K$, and we perform it \emph{once} at the end (i.e., no intermediate collapses). This corresponds to the modeling choice that an attacker can leverage any foothold service at the source node and, on a hop, can attempt all destination interfaces; independence across attempts is captured by the noisy–OR operator $\oplus$. Since $P^\uparrow_K$ includes $I^\uparrow$, node diagonals of $P_K$ are handled automatically.

\begin{figure}[H]
	\centering
	\hspace{-3cm}
	\begin{minipage}{0.58\linewidth}
		\centering
		\begin{tikzpicture}[>=Stealth, scale=0.55]
			\coordinate (i) at (0,0);
			\coordinate (j) at (2.6,1.2);
			\coordinate (k) at (2.6,-1.2);
			\coordinate (l) at (5.2,0);

			\node (ni) [draw, circle, inner sep=1.6pt] at (i) {$i$};
			\node (nj) [draw, circle, inner sep=1.6pt] at (j) {$j$};
			\node (nk) [draw, circle, inner sep=1.6pt] at (k) {$k$};
			\node (nl) [draw, circle, inner sep=1.6pt] at (l) {$\ell$};

			\draw[->] (ni) edge[loop above] node {$1$} (ni);
			\draw[->] (nj) edge[loop above] node {$1$} (nj);
			\draw[->] (nk) edge[loop below] node {$1$} (nk);
			\draw[->] (nl) edge[loop right] node {$1$} (nl);

			\draw[->,bend left=12]  (ni) to node[pos=0.55, above] {$0.86$} (nj);
			\draw[->,bend right=12] (ni) to node[pos=0.55, below] {$0.86$} (nk);
			\draw[->,bend left=12]  (nj) to node[pos=0.55, above] {$0.8$} (nk);
			\draw[->,bend left=14]  (nj) to node[pos=0.55, above] {$0.8$} (nl);
			\draw[->,bend right=14] (nk) to node[pos=0.55, below] {$0.3$} (nl);
		\end{tikzpicture}
	\end{minipage}\hspace{3mm}
	\begin{minipage}{0.30\linewidth}
		{\setlength{\abovedisplayskip}{6pt}%
			\setlength{\belowdisplayskip}{6pt}%
			\setlength{\arraycolsep}{3pt}%
			\[
				P_{1} \;=\;
				\begin{bmatrix}
					1.00 & 0.86 & 0.86 & 0.00   \\
					0.00 & 1.00 & 0.80 & 0.80 \\
					0.00 & 0.00 & 1.00 & 0.30 \\
					0.00 & 0.00 & 0.00 & 1.00
				\end{bmatrix}
			\]
		}
	\end{minipage}
	\caption{Node-level one-hop reach \(P_1\) (identity + one-hop).}
	\label{fig:P1-node}
\end{figure}

\begin{figure}[H]
	\centering
	\hspace{-3cm}
	\begin{minipage}{0.58\linewidth}
		\centering
		\begin{tikzpicture}[>=Stealth, scale=0.55]
			\tikzstyle{v}=[rectangle,draw,rounded corners,minimum width=22pt,minimum height=14pt,inner sep=2pt]
			\coordinate (iS) at (0,0);
			\coordinate (iM) at (0,-4.0);
			\coordinate (jS) at (6,0);
			\coordinate (kM) at (6,-4.0);
			\coordinate (lS) at (10,-2.0);

			\node[v] (iSnode) at (iS) {$(i,\mathrm{SSH})$};
			\node[v] (iMnode) at (iM) {$(i,\mathrm{MSSQL})$};
			\node[v] (jSnode) at (jS) {$(j,\mathrm{SSH})$};
			\node[v] (kMnode) at (kM) {$(k,\mathrm{MSSQL})$};
			\node[v] (lSnode) at (lS) {$(\ell,\mathrm{SSH})$};

			\draw[->,bend left=12]  (iSnode) to node[pos=0.55, above] {$0.64$} (kMnode);
			\draw[->,bend left=6]  (iSnode) to node[pos=0.55, below] {$0.73$} (lSnode);

			\draw[->,bend right=12] (iMnode) to node[pos=0.55, below] {$0.24$} (kMnode);
			\draw[->,bend left=6] (iMnode) to node[pos=0.55, above] {$0.31$} (lSnode);

			\draw[->,bend left=12]  (jSnode) to node[pos=0.55, above] {$0.24$} (lSnode);
		\end{tikzpicture}
	\end{minipage}\hspace{3mm}
	\begin{minipage}{0.30\linewidth}
		{\setlength{\abovedisplayskip}{6pt}%
			\setlength{\belowdisplayskip}{6pt}%
			\setlength{\arraycolsep}{3pt}%
			\[
				W^\uparrow_{2} \;=\;
				\begin{bmatrix}
					0.00 & 0.00 & 0.00 & 0.64 & 0.73 \\
					0.00 & 0.00 & 0.00 & 0.24 & 0.31 \\
					0.00 & 0.00 & 0.00 & 0.00 & 0.24 \\
					0.00 & 0.00 & 0.00 & 0.00 & 0.00 \\
					0.00 & 0.00 & 0.00 & 0.00 & 0.00
				\end{bmatrix}
			\]
		}
	\end{minipage}
	\caption{Lifted exactly-two-hop transition \(W^\uparrow_2 = W^\uparrow \otimes W^\uparrow\).}
	\label{fig:W2-up}
\end{figure}

\begin{figure}[H]
	\centering
	\hspace{-3cm}
	\begin{minipage}{0.58\linewidth}
		\centering
		\begin{tikzpicture}[>=Stealth, scale=0.55]
			\tikzstyle{v}=[rectangle,draw,rounded corners,minimum width=22pt,minimum height=14pt,inner sep=2pt]
			\coordinate (iS) at (0,0);
			\coordinate (iM) at (0,-4.0);
			\coordinate (jS) at (6,0);
			\coordinate (kM) at (6,-4.0);
			\coordinate (lS) at (10,-2.0);

			\node[v] (iSnode) at (iS) {$(i,\mathrm{SSH})$};
			\node[v] (iMnode) at (iM) {$(i,\mathrm{MSSQL})$};
			\node[v] (jSnode) at (jS) {$(j,\mathrm{SSH})$};
			\node[v] (kMnode) at (kM) {$(k,\mathrm{MSSQL})$};
			\node[v] (lSnode) at (lS) {$(\ell,\mathrm{SSH})$};

			\draw[->] (iSnode) edge[loop above] node {$1$} (iSnode);
			\draw[->] (iMnode) edge[loop below] node {$1$} (iMnode);
			\draw[->] (jSnode) edge[loop above] node {$1$} (jSnode);
			\draw[->] (kMnode) edge[loop below] node {$1$} (kMnode);
			\draw[->] (lSnode) edge[loop right] node {$1$} (lSnode);

			\draw[->,bend left=12]  (iSnode) to node[pos=0.55, above] {$0.8$} (jSnode);
			\draw[->,bend right=12] (iSnode) to node[pos=0.55, below] {$0.93$} (kMnode); 
			\draw[->,bend left=6]  (iSnode) to node[pos=0.55, below] {$0.73$} (lSnode);

			\draw[->,bend left=12]  (iMnode) to node[pos=0.55, above] {$0.3$} (jSnode);
			\draw[->,bend right=12] (iMnode) to node[pos=0.55, below] {$0.47$} (kMnode); 
			\draw[->,bend right=6] (iMnode) to node[pos=0.55, above] {$0.31$} (lSnode);

			\draw[->,bend left=12]  (jSnode) to node[pos=0.55, above] {$0.8$} (kMnode);
			\draw[->,bend left=12]  (jSnode) to node[pos=0.55, above] {$0.85$} (lSnode); 

			\draw[->,bend right=12] (kMnode) to node[pos=0.55, below] {$0.3$} (lSnode);
		\end{tikzpicture}
	\end{minipage}\hspace{3mm}
	\begin{minipage}{0.30\linewidth}
		{\setlength{\abovedisplayskip}{6pt}%
			\setlength{\belowdisplayskip}{6pt}%
			\setlength{\arraycolsep}{3pt}%
			\[
				P^\uparrow_{2} \;=\; P^\uparrow_{1} \oplus W^\uparrow_{2} \;=\;
				\begin{bmatrix}
					1.00 & 0.00 & 0.80 & 0.93 & 0.73 \\
					0.00 & 1.00 & 0.30 & 0.47 & 0.31 \\
					0.00 & 0.00 & 1.00 & 0.80 & 0.85 \\
					0.00 & 0.00 & 0.00 & 1.00 & 0.30 \\
					0.00 & 0.00 & 0.00 & 0.00 & 1.00
				\end{bmatrix}
			\]
		}
	\end{minipage}
	\caption{Lifted reach \(P^\uparrow_2\) (identity + one-hop + two-hop union).}
	\label{fig:P2-up}
\end{figure}

\begin{figure}[H]
	\centering
	\begin{minipage}{0.58\linewidth}
		\centering
		\begin{tikzpicture}[>=Stealth, scale=0.55]
			\coordinate (i) at (0,0);
			\coordinate (j) at (2.6,1.2);
			\coordinate (k) at (2.6,-1.2);
			\coordinate (l) at (5.2,0);

			\node (ni) [draw, circle, inner sep=1.6pt] at (i) {$i$};
			\node (nj) [draw, circle, inner sep=1.6pt] at (j) {$j$};
			\node (nk) [draw, circle, inner sep=1.6pt] at (k) {$k$};
			\node (nl) [draw, circle, inner sep=1.6pt] at (l) {$\ell$};

			\draw[->] (ni) edge[loop above] node {$1$} (ni);
			\draw[->] (nj) edge[loop above] node {$1$} (nj);
			\draw[->] (nk) edge[loop below] node {$1$} (nk);
			\draw[->] (nl) edge[loop right] node {$1$} (nl);

			\draw[->,bend left=12]  (ni) to node[pos=0.55, above] {$0.86$} (nj);
			\draw[->,bend right=12] (ni) to node[pos=0.55, below] {$0.96$} (nk);
			\draw[->,bend left=12]  (nj) to node[pos=0.55, right] {$0.8$} (nk);
			\draw[->,bend left=14]  (nj) to node[pos=0.55, above] {$0.85$} (nl);
			\draw[->,bend right=14] (nk) to node[pos=0.55, below] {$0.3$} (nl);
			\draw[->,bend left=12]  (ni) to node[pos=0.35, above] {$0.81$} (nl);
		\end{tikzpicture}
	\end{minipage}\hspace{3mm}
	\begin{minipage}{0.32\linewidth}
		{\setlength{\abovedisplayskip}{6pt}%
			\setlength{\belowdisplayskip}{6pt}%
			\setlength{\arraycolsep}{3pt}%
			\[
				P_{2} \;=\;
				\begin{bmatrix}
					1.00 & 0.86 & 0.96 & 0.81 \\
					0.00 & 1.00 & 0.80 & 0.85 \\
					0.00 & 0.00 & 1.00 & 0.30 \\
					0.00 & 0.00 & 0.00 & 1.00
				\end{bmatrix}
			\]
		}
	\end{minipage}
	\caption{Node-level reach \(P_2\) (identity + union of all 1–2 hop paths via lifted space).}
	\label{fig:P2-node}
\end{figure}

\paragraph{What this shows.}
We use a single model: node splitting by service with land–then–pivot semantics. Hop success depends on the \emph{landing service} carried by the source state, so all compositions are done in the lifted space and we collapse back to nodes only once via a noisy–OR over both source and destination service labels. This preserves landing–service memory across hops; any intermediate collapse to nodes would lose that memory and undercount reach. The resulting $P_K$ feeds directly into $\mathrm{LMS}_K$ and $\mathrm{BRE}_K$.

\subsection{Computational Cost}\label{subsec:computational-cost}
Each iteration of the algorithm is dominated by the probabilistic composition product ($\otimes$), which has a computational complexity of $O(n^3)$, equivalent to standard dense matrix multiplication. The operations involved (primarily multiplications and additions) are highly parallelizable and operate on dense matrices. This makes the algorithm well-suited for hardware acceleration on platforms such as GPUs, ensuring its practical applicability to real-world networks of significant size.

\section{Examples and Evidence}\label{sec:examples}

\subsection{Evidence of Practical Utility}\label{subsec:utility}
We reuse the running 4-asset graph for continuity:
\(i=\) User Workstation, \(j=\) Web/App Server, \(k=\) Mid-tier Service Host, \(\ell=\) Domain Controller.
Binary reachability would mark both \(i{\to}j{\to}\ell\) (two hops) and any direct admin path as equally “reachable.”
Our probabilistic model—via \(P_2\) (Fig.~\ref{fig:prob-P2})—quantifies the asymmetry:
high-privilege remote-admin edges dominate over multi-hop, low-privilege routes, which explains the uplift in
BRE$_2$ (and LMS$_2$) and prioritizes the single most important control.

\subsection{Illustrative Field Example: Service-Class Contrast in an Enterprise Snapshot}
\label{subsec:field-example}
We complement our formal results with an illustrative observation from a production environment described by \citet{tyagi-2025-graph-lateral}. The analysis compares adjacency induced by two services: \emph{RDP} (TCP/3389), an interactive remote-admin channel, and \emph{Epic Cache} (TCP/6966), an application-specific service. The induced RDP adjacency was reported as \emph{nearly complete} (most node pairs reachable directly), whereas the Epic Cache adjacency was \emph{nearly empty} (few pairs reachable). Under our framework, this service-class asymmetry implies substantially higher step-success and union mass for RDP-driven paths, elevating both LMS$_K$ (off-diagonal mean) and BRE$_K$ (mean including the diagonal) relative to app-only services.
The same field note underscores that adversaries frequently achieve lateral movement using \emph{valid accounts} over remote services (RDP/SSH/SMB/WinRM) without relying on unpatched CVEs, aligning with ATT\&CK’s “Remote Services” technique and our CVE-agnostic design. We emphasize that this is a single-environment qualitative example (not a statistical study). Nevertheless, the observation is consistent with our model’s predictions and with practitioner guidance prioritizing constraints on interactive remote services. \citep{tyagi-2025-graph-lateral,mitre-t1021,mitre-ta0008,mitre-t1078}

\section{Conclusion and Future Work}\label{sec:conclusion}
\paragraph{Conclusion.}
We presented a probabilistic, service-aware model for lateral-movement exposure that cleanly separates structure (adjacency \(A\)) from step success (pivot potential \(W\)), composes hops via \(\otimes\), and aggregates paths via \(\oplus\). The running example and field snapshot illustrate how the model prioritizes high-impact controls beyond binary reachability.

\paragraph{Future work.}
\begin{itemize}
	\item \textbf{Technique/dependency modeling:} relax noisy-OR independence by conditioning on attack techniques and shared prerequisites (credentials, agent presence); explore dependency-aware unions.
	\item \textbf{Modern environments:} extend to containers, APIs, and service-mesh edges (ephemeral topologies, short-lived identities).
	\item \textbf{Vuln/RCE coupling:} optional integration of vulnerability evidence (e.g., exploitable RCEs) as \emph{multipliers} on \(W\) while keeping the baseline CVE-agnostic design.
	\item \textbf{Business value weighting:} incorporate asset criticality and blast impact (risk \(=\) probability \(\times\) loss), with weighted BRE/LMS for decision support.
	\item \textbf{Policy optimization:} compute minimal-control sets (segmentation/identity constraints) that maximally reduce \(\mathrm{BRE}_K\) under operational constraints.
\end{itemize}


\end{document}